
\magnification=\magstep1
\def\nopage0{\pageno=0 \footline={\ifnum\pageno<1 {\hfil} \else
              {\hss\tenrm\folio\hss}   \fi}}
%

%
\def\sqr#1#2{{\vcenter{\hrule height.#2pt
      \hbox{\vrule width.#2pt height#1pt \kern#1pt
         \vrule width.#2pt}
      \hrule height.#2pt}}}
\def\square{{\mathchoice {\ \sqr44\ } {\ \sqr44\ } {\ \sqr34\ } {\ \sqr34\ }}}
%
\def \d2dots{\mathinner{\mkern1mu\raise1pt\vbox{\kern7pt\hbox{.}}\mkern2mu
\raise4pt\hbox{.}\mkern2mu\raise7pt\hbox{.}\mkern1mu}}
%
\def\smdrct#1{{\vcenter{\hbox{\vrule width.4pt height#1pt}}\kern-1.5pt\times}}

%

%

\def\cG{{\cal G}}

                    \def\cR{{\cal R}}
                    \def\cU{{\cal U}}
\def\cV{{\cal V}}                    
          
%

\def\CC{\rlap {\raise 0.4ex \hbox{$\scriptscriptstyle |$}}
\hskip -0.1em C}
\def\FF{\hbox to 8.33887pt{\rm I\hskip-1.8pt F}}
\def\NN{\hbox to 9.3111pt{\rm I\hskip-1.8pt N}}
\def\PP{\hbox to 8.61664pt{\rm I\hskip-1.8pt P}}
\def\QQ{\rlap {\raise 0.4ex \hbox{$\scriptscriptstyle |$}}
{\hskip -0.1em Q}}
\def\RR{\hbox to 9.1722pt{\rm I\hskip-1.8pt R}}
\def\ZZ{\hbox to 8.2222pt{\rm Z\hskip-4pt \rm Z}} 
\nopage0
\line{\hfil CERN-TH.6324/91}
\vfil
\centerline {\bf New fusion rules and $\cR$-matrices for $SL(N)_q$ at roots
of unity} \vfil
\centerline {Daniel Arnaudon
\footnote{*}{On leave from Ecole Polytechnique, 91128 Palaiseau, FRANCE}}
\medskip
\centerline {Theory Division, CERN}
\centerline {CH-1211 Gen\`eve 23, Switzerland}
\vfil
\vfil
\centerline {\bf Abstract}
\medskip
We derive fusion rules for the composition of $q$-deformed classical
representations  (arising in tensor products of the fundamental
representation) with semi-periodic representations of $SL(N)_q$ at roots of
unity. We obtain full reducibility into semi-periodic representations.
On the other hand, heterogeneous
$\cR$-matrices which intertwine   tensor products of periodic or
semi-periodic
representations with $q$-deformed classical
representations   are given. These $\cR$-matrices satisfy  all the possible
Yang Baxter equations with one another and, when they exist, with the
$\cR$-matrices intertwining homogeneous  tensor products of periodic or
semi-periodic
representations.  This compatibility between these two  kinds of
representations has never been used in physical models.
\vfil \line{
CERN-TH.6324/91 \hfil} \line{ November 1991 \hfil}
\eject
Quantum groups [Dri, Jim, F.R.T] at roots of unity [Lus, D-C.K, D-C.K.P]
now play an important role in
physics.
When the deformation parameter $q$ of the
quantum group  is a root of unity, we can distinguish two kinds of
irreducible  representations (irreps):
\item{--}{ Type A irreps:
the so-called $q$-deformed representations. They are the deformation of
representations that exist for $q=1$, and have the same structure. They are
highest-weight and lowest-weight representations, and do not depend on
supplementary complex continuous parameters. }
\item{--}{ Type B irreps:
those irreps which do not match the previous
definition. These representations have been studied and almost classified
(See [Skl, R.A] for $SL(2)_q$ and [A.C, D-C.K, D-C.K.P, D.J.M.M.2, Dob]).
Their dimension is bounded and they are described by a set
of continuous complex parameters corresponding to the eigenvalues of the
generators of the augmented centre. The generators can be represented
in particular
by invertible matrices. We will call periodic irreps those for
which all the generators related to positive and negative roots are
represented by invertible matrices, and semi-periodic the lowest-weight
irreps for
which the generators related to positive roots only are
represented by invertible matrices.}

\medskip
The first family appears in conformal field theories
(see e.g. [A-G.G.S, M.R])  as well as in
statistical physics of integrable models [Pas, P.S].
The set of type A representations is
not stable under fusion (tensor product). The fusion rules for these
representations involve indecomposable representations [P.S, Kel]. However,
these fusion rules are generally truncated for physical purposes
[A-G.G.S, F.G.P].

The  type  B representations
are used in physics in relation with integrable models. They  first
appeared in the study of the eight-vertex model with the Sklyanin
algebra
[Skl], whose trigonometric limit is $SL(2)_q$.
They are now used in relation with
the so-called generalized chiral
Potts model [B.S, D.J.M.M, K.M.S] (periodic
irreps)  and with relativistic
solitons [G.S] (semi-periodic irreps).
(See e.g. [G.R, A.MC.P] about the chiral Potts model.)
In [D.J.M.M], the fusion rules of
minimal periodic representations of $SL(N)$ (type B) are considered.
The tensor
products of such representations is actually studied as a representation of
$\widehat{SL(N)}_q$ and is proved to be irreducible under the action of this
quantum algebra. In [G.R-A.S, G.S], the fusion rules for
semi-periodic irreps of $SL(2)$ are studied.

\medskip
When $q$ is not a root of unity, there exists a universal $\cR$-matrix,
satisfying the Yang Baxter equation (see
[Ros] for an expression on the case of $SL(N)_q$).
When
one evaluates this universal $\cR$-matrix on tensor products of
representations,
one get intertwiners, i.e. invertible matrices that express the equivalence
of  differently ordered tensor products.
When now $q$ is a root of unity, the
universal $\cR$-matrix diverges. Formal evaluation of it on irreps of type A
still provides intertwiners for these representations.
However, differently-ordered tensor products of type B irreps are not always
equivalent  [D.J.M.M, G.R-A.S, G.S].
When they are equivalent, the intertwiner is not
related to the ill-defined universal $\cR$-matrix. However it can still be
used to give the Boltzmann weights of some statistical models.

Although both families of irreducible representations appear in physics,
they have never been used together in
the same physical model. The $q$-deformed representations (type A) can
appear in some
degenerate limits of the parameters of representations of the second family
(B). But the  $\cR$-matrix of models based on family B is not well defined
in such limits
\footnote{*}{The author thanks C. G\'omez for a discussion on that point.}.
\medskip
In this letter, we
consider the $SL(N)_q$ case.
We first
give new fusion rules corresponding to
the composition of type A irreps with only semi-periodic (type B) irreps.
On the other hand, we define
$\cR$-matrices acting on heterogeneous tensor products
involving both types A and B of representations (all irreps of type B
included). This letter is a generalization of [Arn] and a step toward the
definition of physical  models involving two kinds of states related with the
two kinds of  irreducible representations. As an application, the construction
of new integrable quantum chains will be sketched in the conclusion.

\bigskip
The quantum group
$SL(N)_{q}$ is defined by the generators
$k_{i}$, $k_{i}^{-1}$, $e_{i}$, and $f_{i}$, for
$i=1,...,N-1$,
and the relations
$$\left\{ \eqalign{
  k_{i} k_{i}^{-1} &= k_{i}^{-1}k_{i} =1\; , \qquad
  k_{i} k_{j} = k_{j}  k_{i} \;  ,
\cr \cr
  k_{i}  e_{j}  k_{i}^{-1} &= q^{a_{ij}} e_{j}   \;  , \qquad
  k_{i}  f_{j}  k_{i}^{-1} = q^{-a_{ij}} f_{j}  \;  ,
\cr \cr
  [e_{i},f_{j}] &= \delta _{ij} {k_{i} - k_{i}^{-1} \over q -q^{-1}}\; ,
\cr \cr
  [{e}_{i},{e}_{j}] &=[{f}_{i},{f}_{j}] =0 \qquad {\rm for} \qquad
            \left|{i-j}\right|\ge 2 \;,
\cr \cr
  e_{i}^{2} e_{i\pm 1}- & (q+q^{-1})  e_{i} e_{i\pm 1} e_{i}
     +e_{i\pm 1} e_{i}^{2}   =0\; ,
\cr \cr
  f_{i}^{2} f_{i\pm 1}- & (q+q^{-1})  f_{i} f_{i\pm 1} f_{i}
     +f_{i\pm 1} f_{i}^{2}   =0\;,
\cr   } \right. \eqno(1)$$
where $(a_{ij})_{ i,j=1,...,N-1}$ is the Cartan matrix of $SL(N)$, i.e.
$$\cases{ a_{ii}=2\cr
         {a}_{ii\pm 1}=-1\cr
         {a}_{ij}=0 & for $\left| i-j \right|\ge 2$ \cr} \eqno(2) $$
The coproduct is defined by
$$\left\{
\eqalign{
&\Delta(e_i) = e_i \otimes 1        + k_i \otimes e_i  \cr
&\Delta(f_i) = f_i \otimes k_i^{-1} + 1   \otimes f_i  \cr
&\Delta(k_i) = k_i \otimes k_i                         \cr
&\Delta ~:~ SL(N)_q \rightarrow SL(N)_q \otimes SL(N)_q \qquad
\hbox{ homomorphism  of  algebras,}                      \cr }
\right.  \eqno(3) $$
while the opposite coproduct $\Delta '$ is $\Delta '= P \Delta P$ where $P$
is the permutation map $P x\otimes y =y \otimes x$. (The conventions here
are slightly different from those of [Arn].)
Denoting by
$\alpha_i\equiv \alpha_{i,i+1}$,
for $i=1,...,N-1 $, the simple roots of $SL(N)$,
all the positive roots can be written
$\alpha_{ij}=\sum _{k=i}^{j-1} \alpha_k $
for $1\le i < j \le N$. We denote by $R_+$ the set of positive roots.
As usual, the root vectors $e_{ij}\equiv e_{\alpha_{ij}}$ are defined
inductively by
$$ e_{ij+1}=e_{ij}e_{jj+1}-qe_{jj+1}e_{ij}\;, \eqno(4) $$
so that for $1\le i<j<k \le N$
$$ e_{ik}=e_{ij}e_{jk}-qe_{jk}e_{ij}\;.\eqno(5) $$
(There are other possible choices for the definition of root vectors,
depending on the choice of the way of writing the longest element of the
Weyl group. This choice is the most standard.)
One can similarly define root vectors $f_{ij}$ associated to negative roots.
\medskip
Let $q$ be a primitive $m'$-th root of unity. Denote by $m$ either $m'$ if
$m'$ is odd, or $m'/2$ if $m'$ is even.
Then, as proved in [D-C.K, D-C.K.P],  the dimension of the irreducible
representations (type A or B) is bounded by $m^{N(N-1)/2}$.
All the elements
$e_{ij}^m$,   $f_{ij}^m$, and $k_i^m$ belong to the centre or $SL(N)_q$, and
the
irreducible representations of type B are labelled by at most
$N^2-1$ complex
continuous parameters corresponding to the values of these
elements of the centre.

\medskip
\noindent
{\bf FUSION RULES:}
\medskip
We will now consider the fusion of semi-periodic representations with the
fundamental representation, and then with any type A irrep.
\medskip
The fundamental representation $\pi_\square$ of $SL(N)_q$ on the vector space
$V_{\square}$ spanned by  $w_j$ ($j=1,...,N $) is given by
$$
\cases{
\pi_\square(
f_i) w_j = \delta_{ij} w_{j+1}     &  $1 \le j \le N-1 $  \cr
\pi_\square(
f_{i}) w_N =0                      &                      \cr
\pi_\square(
e_i) w_j = \delta_{i,j-1} w_{j-1}  &  $2 \le j \le N $    \cr
\pi_\square(
e_i) w_1 = 0                       &                      \cr
\pi_\square(
k_i) w_j = q^{ \delta_{ij}-\delta_{i,j-1} } w_j
                                   &  $1 \le j \le N $    \cr }
\eqno(6) $$
\medskip
A generic semi-periodic
representation $\pi_x$ on the vector space $V$ is defined
as follows: it has the maximal dimension  $m^{N(N-1)/2}$ allowed for
irreducible representations when $q^{m'}=1$. The central elements $e_{ij}^m$
take the value $x_{ij}\in \CC\backslash\{ 0 \}$ on $V$.
It is a lowest-weight representation, i.e. it contains a
vector $v_0$ such that for all $i\in \{1,...,N-1\}$, $f_i v_0 =0$.
This vector is
a common eigenvector  of the $k_i$'s and
$k_i v_0=\lambda_i v_0 \equiv q^{\mu_i} v_0$,
so that the
central elements $k_i^m$ take the value $z_i=\lambda_i ^m = q^{m\mu_i }$.
The parameters characterizing this representation are then $x_{ij}$
($1\le i < j \le N-1$) and $\lambda_i$  ($1\le i \le N-1$). Their number
$N(N-1)/2+N-1$ is less than the maximum since the parameters corresponding to
the eigenvalues of  $f_{ij}^m$ vanish.
For generic value of the parameters
$x\equiv (x_{ij},\lambda_i)$, this representation $\pi_x$ is irreducible, and
$v_0$ is the only singular vector [D-C.K]
\footnote{*}{Note however that these two properties are not
equivalent. There are actually representations with other singular vectors
(for non-generic values of the parameters) which are still
irreducible.}.
The condition for this (irreducibility and uniqueness of the singular vector)
is that none of the $x_{ij}$'s vanishes, and that  none of the products
$\lambda_i ^2\lambda _{i+1} ^2... \lambda_{j -1}^2$ ($i<j$) is a power of
$q$.

Using an analogue of the P.B.W. basis of $\cU_q (\cG ^+)$ [Ros], a basis
of this representation is given by [D-C.K],
$$v_p=\pi_x(e^p) v_0 \equiv \pi_x \left(
                     e_{N-1,N}^{p_{N-1,N}}\
                     e_{N-2,N}^{p_{N-2,N}}\
                     e_{N-2,N-1}^{p_{N-2,N-1}}\ ...\
                     e_{2,3}^{p_{2,3}}\
                     e_{1,N}^{p_{1,N}}\ ...\
                     e_{1,3}^{p_{1,3}}\
                     e_{1,2}^{p_{1,2}}\   \right) v_0 \eqno(7) $$
where $p=(p_{ij})_{1\le i <j\le N-1}$
and $p_{ij} \in \{0,..,m-1\}$. (The $e_{ij}$ are written in the
inverse lexical order.)
\medskip
This representation is precisely given by:
$$ v_{p_{ik}+m} \equiv x_{ik} v_p,$$
and the actions of the generators of $SL(N)$
on $v_p$
$$\eqalign{
\pi_x(k_i) v_p = &
q^{\mu_i +\sum\limits_{\alpha '}p_{\alpha '}(\alpha ',\alpha_i)}  v_p
\;, \cr
\pi_x(e_i) v_p = &
q^{-\sum_{\alpha>\alpha_i} p_{\alpha }(\alpha ,\alpha_i)} v_{p_{i,i+1} +1}
+
\sum_{k=i+2}^N [p_{i+1, k}] q^{\sum_{k'>k} (p_{i+1, k'} - p_{ik'})  }
v_{p_{i+1 k}-1,p_{ik}+1}
\;, \cr
\pi_x(f_i) v_p = &\sum _{k>i+1} [p_{ik}]
q^{\mu_i -\sum \limits _{i+1<l<k}p_{i+1,l}+\sum \limits
_{\alpha '<\alpha_{ik}} p_{\alpha '}(\alpha_i,\alpha ')}
v_{p_{i+1,k}+1,p_{ik}-1}
\cr
& \qquad
- \  [p_{i,i+1}] \left[\mu_i+p_{i,i+1}-1+\sum _{\alpha '<\alpha_{i.i+1}}
p_{\alpha '} (\alpha_i,\alpha ') \right] v_{p_{i,i+1}-1}
\cr
& \qquad - \ \sum_{j<i}  [p_{j,i+1}]
q^{-\mu_i +2 -\sum \limits _{\alpha '\le \alpha_{j,i+1}}
p_{\alpha '} (\alpha_i,\alpha ')} v_{p_{j,i+1}-1,p_{ji}+1}
\cr
}\eqno(8) $$
where only the modified indices of $v$ are written  on the  right-hand side,
and where the symbols $<$ and $\le$ refer to the lexical order of the roots.
$(,)$ denotes the bilinear form defined on the root lattice by
$(\alpha_i,\alpha_j)=a_{ij}$.
(As usual, $[a]=(q^a-q^{-a})/(q-q^{-1})$.)

\medskip
{\bf Proposition: }{\sl  The tensor product
$\left( V_x \otimes V_{\square} ,
(\pi_x \otimes \pi_{\square}) \circ \Delta \right)$
of
an irreducible semi-periodic representation $\pi_x $
having a single singular (=lowest-weight) vector,
by the fundamental representation
is totally reducible  and
$$V_{(x_{ij},\lambda_i)}  \otimes V_{\square} =
\bigoplus _{n=0}^{N-1}
V_{(x_{ij},\lambda_i^{(n)}=\lambda_i q^{\delta_{i,n+1}-\delta_{i,n}})}
\eqno(9) $$}
\medskip
{\it Proof:}
\item {a) }{ Although the expression of $\Delta(e_{ij})^p$ on the  tensor
product is quite complicated [Ros], it becomes very simple on $V_x \otimes
V_{\square}$  for $p=m $.  Actually,
$$(\pi_x \otimes \pi_{\square}) ( \Delta(e_{ij}))^m =
\pi_x (e_{ij}^m) \otimes 1 = x_{ij}\;.
\eqno(10) $$}
\item {b) }{ One can find in $V_x \otimes V_{\square}$ exactly $N$ lowest
weight vectors $v_0 ^{(n)}$ ($n=0,...,N-1$). Explicitly, these vectors read
$$v_0 ^{(n)} = \sum _{l=0}^{n} \cV _l ^{(n)} \otimes w_{N-n+l}\;.
\eqno(11) $$
They satisfy
$$(\pi_x \otimes \pi_{\square})  \Delta(f_i) v_0 ^{(n)}=0\eqno(12) $$
for $i=1,...,N-1$,
provided
$$\cases{
\cV_l ^{(n)} = - q \pi_x(f_{N-n+l}) \cV_{l+1} ^{(n)}
& for $0\le l <n \le N-1$     \cr
\pi_x (f_i) \cV_n ^{(n)} = 0
& for $1\le i \le n-2$        \cr}
\eqno(13) $$
Explicitly, $\cV_n ^{(n)}= \sum_p a_p v_p$,   where the
sum is limited to $v_p$ of the type
$$e_{i_R N}e_{i_{R-1}i_R}...e_{i_{r}i_{r+1}}...e_{i_1 i_2} v_0$$
with
$ N-n = i_1 < i_2 <...< i_R < N $ and
$$ a_p = \prod _{l\in \{N-n,...,N-1\}\backslash \{i_1,...,i_R\}}
{1-q^{-2\sum\limits_{l'=l-1}^{N-n}(\mu_{l'}-1) -2} \over q-q^{-1}}
\eqno(14) $$}
\item{c) }{
Let
$V_{(x_{ij},\lambda_i^{(n)}=\lambda_i q^{\delta_{i,n+1}-\delta_{i,n}})}$
(for $n=0,...,N-1$)
be
the semi-periodic representations generated by the action of
$(\pi_x \otimes \pi_{\square}) ( \Delta (SL(N)_q))$ on
$v_0 ^{(n)}$. They are irreducible and have a single lowest-weight vector.
It is
then easy to prove that the sum of
these vector spaces is a direct sum, so that the proposition is proved.}
\medskip
{\bf Corollary:} {\sl
\item{a) }{
The tensor product of a generic semi-periodic representation with any
irreducible representation $(V_J,\pi_J)$ of type A is totally
reducible and
$$V_{(x_{ij},\lambda_i)}  \otimes V_J =
\bigoplus _{\mu ' \hbox{ weight of } \pi_J  } m(\mu ')
V_{(x_{ij},\lambda_i^{(n)}=\lambda_i q^{\mu _i'})},
\eqno(15) $$
where $m(\mu ')$ denotes the multiplicity of the weight $\mu '$ of $\pi_J$.}
\item{b) }{
The same holds for the tensor product
of a generic semi-periodic representation with an indecomposable
representation that can appear in the composition of type A irreps.} }
\medskip
This can be proved directly as in [Arn] for $N=2$ (by finding all the
singular vectors)
or it can also be seen as a consequence of the co-associativity
of $\Delta$.
\medskip
What happens now for the composition of periodic representations (with no
lowest weight) with ordinary representations~? In the case $N=2$, one can
prove [Arn2] that the quadratic Casimir takes generically two values on the
tensor product of a periodic representation with the fundamental
representation. So the full reducibility stated in the proposition (and its
corollary) extends to periodic representations (except for non-generic values
of the parameters), in the case of $SL(2)_q$. The case $N>2$ is not treated
yet.
\medskip
Let us end this part devoted to fusion rules with some remarks on the
composition of type B irreps with type B irreps in the case of $SL(2)_q$.
As proved in
[G.R-A.S, G.S], the tensor product of two semi-periodic
representations is reducible into semi-periodic representations. Looking again
at the quadratic Casimir ($C_2$), we can extend this result (reducibility) to
periodic representations [Arn2]: some parameters of the representations enter
indeed in the characteristic polynomial of $C_2$ only in the constant term,
which proves that for generic values of the parameters the roots of this
polynomial are different.  (This reducibility holds with respect to $SL(2)_q$.
Extensions of these
representations to representations    of  $\widehat{SL(2)}_q$
generate, under tensor products, {\it irreducible}
representations with respect
to  $\widehat{SL(2)}_q$, as already known [D.J.M.M].)

\bigskip
\noindent
{\bf $\cR$-MATRICES:}
\medskip
Let us now consider the problem of $\cR$-matrices for the tensor product of
type A and type B representations.
We look for an $\cR$-matrix $\cR (x,J)$
intertwining  $V_x \otimes V_J$ and $V_J \otimes V_x$, i.e. satisfying
$$ \forall X \in SL(N)_q  \qquad
\cR  (x,J)(\pi_x \otimes \pi_J) \circ \Delta (X)
=  (\pi_x \otimes \pi_J) \circ \Delta ' (X) \cR (x,J)\;,
\eqno(16) $$
where $(V_x,\pi_x)$ denotes a type B irrep.
($\cR$ does not contain the
permutation map with this convention, and the intertwiner is actually $P\cR$.)

A solution of (16) is
the evaluation $(\pi_x \otimes \pi_J)(\cR_u)$ of the truncated
universal
$\cR$-matrix
$$ \cR_u= q^{-b_{ij}h_i \otimes h_j }
\prod_{\alpha \in R_+} \left(
\sum_{n=0}^{m-1} q^n {(1-q^2)^n \over [n]! } q^{-n(n-1)/2}
(k_{\alpha}^{-1} e_{\alpha })^n \otimes  (k_{\alpha} f_{\alpha })^n \right)
\eqno(17) $$
where the matrix $(b_{ij})$ is the inverse of the Cartan matrix, and
where the order of the product is given by the lexical order of the positive
roots.
Let us denote by $\cR^+ (x,J)$ this solution.
We define similarly $\cR^+ (J,x)$.
Let us also denote by $\cR^+ (J,J')$
the evaluation $(\pi_J \otimes \pi_{J'})(\cR_u )$, which intertwines
$\Delta$ and $\Delta ' $ on $V_J \otimes V_{J'} $.

There is another solution to (16), given by the evaluation of the
truncated, permuted inverse of the universal  $\cR$-matrix
$$ \tilde \cR_u= q^{b_{ij}h_i \otimes h_j }
\prod_{\alpha \in R_+} \left(
\sum_{n=0}^{m-1} {(q-q^{-1})^n \over [n]! } q^{n(n-1)/2}
\tilde f_{\alpha }^n \otimes  \tilde e_{\alpha }^n \right)
\eqno(18) $$
where the $\tilde e_{\alpha}$ are the root vectors related with the
reverse order of
the roots, i.e.
$$\cases{
\tilde e_{\alpha_i} = e_i \cr
\tilde e_{ik}=\tilde e_{ij}\tilde e_{jk}-q^{-1} \tilde e_{jk}\tilde e_{ij} \cr
}\eqno(19) $$
($\tilde f_{\alpha } $ is defined similarly.)
$\tilde \cR_u$ is equal to $ P( \cR_u ^{-1})$ modulo powers of the generators
higher or equal to $m$.
Let us denote  $\cR^- (x,J)=(\pi_x \otimes \pi_J) (\tilde \cR_u) $.
We define similarly $\cR^- (J,x)$ and $\cR^- (J,J')$.

\medskip
{\bf Theorem:} {\sl
Let $(V_x,\pi_x)$, $(V_{x'} ,\pi_{x'} )$ be two representations for which
there exists an intertwiner $\cR (x,{x'} )$, and
$(V_{J},\pi_{J})$ a type A irrep. Then the following
Yang Baxter equations are satisfied,
\item{a) } {
On $V_x \otimes V_{x'}  \otimes V_J$,
$$\cR _{12}(x,{x'} )\cR^+  _{13}(x,J)\cR^+  _{23}({x'} ,J)
=\cR^+  _{23}({x'} ,J)\cR^+  _{13}(x,J)\cR _{12}(x,{x'} )\;.\eqno(20) $$
}
\item{b) } {
On $V_x \otimes V_J \otimes V_{x'}$,
$$\cR^+  _{12}(x,J )\cR _{13}(x,x')\cR^-  _{23}(J ,x')
=\cR^-  _{23}(J ,x')\cR _{13}(x,x')\cR^+  _{12}(x,J )\;.\eqno(21) $$
}
\item{c) } {
On $V_{J}  \otimes V_x \otimes V_{x'}$,
$$\cR^-  _{12}(J,x )\cR^-  _{13}(J,x')\cR _{23}(x ,x')
=\cR _{23}(x ,x')\cR^-  _{13}(J,x')\cR^-  _{12}(J,x)\;.
\eqno(22) $$
}
\item{d) } { One can replace in a), b) and c) above one or both of the type B
representations $(V_x,\pi_x)$ and $(V_{x'} ,\pi_{x'} )$ by type A irreps,
changing $\cR (x,{x'} )$ to the
corresponding $\cR^+ $ (or also $\cR^- $), and the eqns. (20-22) are still
valid. Furthermore, all the type A irreps can also be replaced by
indecomposable representations occuring in the fusion rules of type A irreps.
Finally,  $\cR^+ $ and $\cR^- $ can be exchanged globally in each equation.}

{\bf However},
\item{e) } { the Yang Baxter equation }
$$\cR^+  _{12}(x,J)\cR _{13}(x,{x'} )\cR^+  _{23}(J,{x'} )
=\cR^+  _{23}(J,{x'} )\cR _{13}(x,{x'} )\cR^+  _{12}(x,J)\eqno(23) $$
\item {}{ {\bf cannot} be satisfied on $V_x \otimes V_J \otimes V_{x'} $
for generic $x$ and $x'$.} }
\medskip
{\it Proof:}
\item{a) } {
follows from the fact that $\cR (x,{x'} )$ is an intertwiner for   $\Delta$
and  $\Delta '$ on $V_x \otimes V_{x'} $. One proves it first with
$J=\square$.
[The expression
$$
\Delta(e_{\alpha}) = e_{\alpha} \otimes 1 + k_{\alpha} \otimes e_{\alpha}
+\sum _{\beta + \gamma = \alpha \atop \beta > \gamma} (1-q^2)
k_{\gamma }e_{\beta } \otimes e_{\gamma }
$$
of the coproduct of $e_{\alpha }$ for any $\alpha\in R_+$
(not necessarily simple) is needed in the computation.]
One then goes to any $J$ using the quasi-triangularity property of
$\cR^+ (x,J)$: if $V_J$ enters in the decomposition of
$V_{J_1} \otimes V_{J_2}=\bigoplus_{J'} V_{J'}$, then
$$\cR^+  (x, J)=({\bf 1} \otimes p) ({\bf 1}
\otimes CG ) \left(
 \cR^+  _{1,2}(x, J_1)\cR^+  _{1,3}(x, J_2) \right)
({\bf 1} \otimes CG^{-1})
({\bf 1} \otimes i)\;,
\eqno(24) $$
where  $p$ is the projector on $V_J$ in the decomposition of
$V_{J_1} \otimes V_{J_2}$, whereas $i$ is the injection of $V_J$ into
$\bigoplus_{J'} V_{J'}$.
$CG$ is the
Clebsch--Gordan invertible maps
$CG \in {\rm End} (V_{J_1} \otimes V_{J_2} , \oplus _{J'} V_{J'})$ such that
$$\forall X\in SL(N)_q, \qquad
p \circ CG\circ
\left( \pi_{J_1}\otimes \pi_{J_2} \right) \left( \Delta (X) \right)
= \pi_J (X) \circ p \circ CG\;. \eqno(25) $$
Note that the
explicit expression for $\cR (x,{x'} )$ is not necessary for the proof. }
\item{b,c,d) } { b, c and d are consequences of a.}
\item{e) } { The constraints provided by the Yang Baxter equation
on  $V_x \otimes V_J \otimes V_{x'} $  with this choice for $\cR (x,J )$
and $\cR (J,{x'} )$ generate
constraints on either $x$ or ${x'} $.}
\medskip

This theorem leads to the following
\medskip
{\bf Corollary:} {\sl  Consider a set of
representations $(V_x,\pi_x)$ such that
each pair of them is intertwined by some $\cR$-matrix (related to chiral
Potts model oranything else), these $\cR$-matrix satisfying all together Yang
Baxter equations. Then one can add to this set all the type A
irreps $(V_J,\pi_J)$ and all the indecomposable representations appearing in
their tensor products. With $\cR^+  (x, J)$,
$\cR^+  (J, J')$ and $\cR^-  (J, x)$, the whole set of
$\cR$-matrices satisfy all the possible Yang Baxter equations. }
\medskip
{\it Remark:} As shown by part e) of the theorem,
the ``canonical'' choice   $\{ \cR^+  (x, J)$, $\cR^+  (J, x) \}$
does not work.
The intertwiner for $\Delta$ and $\Delta '$ on $V_J \otimes V_x$ has to be
the inverse of the one on $V_J \otimes V_x$.
This ``triangularity'' property  is to be compared to the
fact that the known solutions [D.J.M.M, G.R-A.S, G.S] for  $\cR (x, x')$
satisfy $\cR (x', x) = \left( \cR (x, x') \right) ^{-1}$.
However, the whole
set of $\cR$-matrices does  not satisfy this property, since it does not apply
to  $\cR^+  (J, J')$.

It was proved in [Arn] in the case of $SL(2)_q$ that $\cR^+ (x,1/2)$,
$\cR^- (x,1/2)$, $\cR^+ (x,1/2)$ and  $\cR^- (x,1/2)$ were the
{\it only}
intertwiners compatible with the Yang Baxter equations involving $\pi_x$ once
and $\pi_{1/2}$ twice and with the constraint that the intertwiner for
$\pi_{1/2} \otimes \pi_{1/2}$ is $\cR^+ (1/2,1/2)$.
A natural conjecture is that
the only solutions for $\cR (x, J)$ and $\cR (J, x)$ of the
equations (20-22) ($\cR (J, J')$  and $\cR (x, x')$ being given) are
precisely
the evaluations on the representations of (17) and (18) and vice versa.

\bigskip
An application of these results will be the possibility of adding physical
states corresponding to representations $(V_J,\pi_J)$, to integrable theories
that do not already involve such states.

Let us for example consider the
generalization of the chiral Potts model, related with minimal periodic
representations of $SL(N)_q$. Date, Jimbo, Miki and Miwa [D.J.M.M] have
related the intertwiner of minimal periodic representations of $SL(N)_q$
with the
Boltzmann weights of the generalized chiral Potts model. Using the
irreducibility (rather than indecomposability)
under $\widehat{SL(N)}_q$ of the tensor products of these
representations, they proved the Yang Baxter equation for this intertwiner.
We now propose to add new states to this model, corresponding to the
non-periodic representations $(V_J,\pi_J)$. Choosing the intertwiners as
explained in the corollary, the integrability of the generalized Potts model
will ensure that of the enlarged one.

One could also consider a generalization to $SL(N)_q$ of the context of
[G.R-A.S, G.S]  (where $SL(2)_q$ was considered)
and add $q$-deformed classical
states to the semi-periodic states of the theory.

\medskip
We would like now to show an example of a one-dimensional quantum chain which
is $SL(2)_q$ invariant. For $q=i$, i.e. $m=2$, one takes
ordinary spins (fundamental
representation)   on odd sites  and periodic representations (also
two-dimensional) on even  sites. Let us write the periodic representation as
follows:  $$\left\{ \matrix{ &f v_0 = v_1
&e v_0 = y^{-1} \beta v_1
&k v_0 = q^\mu v_0           \cr
&f v_1 = y v_0
&e v_1 = ([\mu ]+\beta) v_0
&k v_1 = -q^\mu v_1          \cr
} \right.
\eqno(26)$$
where $\mu$, $y$, and $\beta$ are three independent complex parameters. The
hamiltonian is
$$H=\sum_{j=1}^{L-1} H_j$$
with
$$ \eqalign{
   H_ {2j}= &
       {1\over 2}q^{\mu}           \sigma_{2j} ^z
     - i ({1\over 2} [\mu] +\beta) \sigma_{2j+1}^z                    \cr
   & + i y q^\mu                   \sigma_{2j}^+ \sigma_{2j+1} ^+
     + y^{-1} \beta                \sigma_{2j}^- \sigma_{2j+1} ^-
     + \left([\mu] +\beta\right)   \sigma_{2j}^+ \sigma_{2j+1} ^-
     - i  q^\mu                    \sigma_{2j}^- \sigma_{2j+1} ^+     \cr}
\eqno(27)
$$
and
$$ \eqalign{
   H_ {2j-1}= &
       {1\over 2}q^{-\mu}          \sigma_{2j} ^z
     + i ({1\over 2} [\mu] +\beta) \sigma_{2j-1}^z                    \cr
   & +   y                         \sigma_{2j-1}^+ \sigma_{2j} ^+
     - i  q^{-\mu} y^{-1} \beta    \sigma_{2j-1}^- \sigma_{2j} ^-
     +                             \sigma_{2j-1}^+ \sigma_{2j} ^-
     + i  q^{-\mu} \left([\mu] +\beta\right) \sigma_{2j-1}^- \sigma_{2j}^+
   \cr}
\eqno (28)
$$
($i=\sqrt{-1}$.)
This chain is trivially integrable by a Jordan--Wigner transformation.
Its interesting feature is the dependence on complex parameters: $q$
is fixed but there are three new parameters. (It is still necessary to check
whether these parameters do not all disappear in equivalence
transformations.) Because of the fusion rules of these two-dimensional
representations into two-dimensional
representations, there will be a zero-mode in the spectrum.
This quantum chain (and other examples) will be studied
elsewhere, following general methods of statistical mechanics [Nijs]. It
will be interesting to compare such chains depending on parameters of
representations with chains based on ordinary representations
of multiparameters quantum groups [H.R].

\bigskip
{\bf Aknowledgements:}
The author wishes to thank D. Altsh\"uler, M. Chaichan, A. Coste,
G. Felder,
A. Ganchev, M. Mintchev, T. Miwa, G. Zemba and particularly
Vladimir Rittenberg  for interesting
discussions and suggestions.

\vskip .5cm
{\bf References: }

\vskip .1cm  \item{[A.C]}{  D. Arnaudon and A. Chakrabarti,
{\sl Periodic and partially periodic representations of $SU(N)_{q}$,}
Commun. Math. Phys. {\bf 139} (1991) 461 and
{\sl Flat periodic representations of
$\cU_{q}(\cG)$,} Commun. Math. Phys. {\bf 139} (1991) 605.}

\vskip .1cm  \item{[A-G.G.S]}{ L. Alvarez-Gaum\'e, C. G\'omez and G. Sierra,
{\sl Hidden quantum symmetries in rational conformal theories,} Nucl.
Phys. {\bf B319} (1989) 155;
{\sl Quantum group interpretation of some conformal field theories,} Phys.
Lett. {\bf B220} (1989) 142;  {\sl Duality and quantum groups,} Nucl. Phys.
{\bf B330} (1990) 347.}

\vskip .1cm  \item{[A.MC.P]}{ G. Albertini, B. McCoy and J. Perk, {\sl
Eigenvalue spectrum of the superintegrable chiral Potts model,} Adv. Stud. in
Pure Math. {\bf 19} (1989) 1-55, and ref. therein.}

\vskip .1cm  \item{[Arn]}{ D. Arnaudon, {\sl Fusion rules and $\cR$-matrix
for the composition of regular spins with semi-periodic representations of
$SL(2)_q$.} Phys. Lett. {\bf B268} (1991) 217.}

\vskip .1cm  \item{[Arn2]}{ D. Arnaudon, {\sl Fusion rules and $\cR$-matrix
for representations of $SL(2)_q$ at roots of unity,} to appear in the
proceedings of the V$^{\rm th}$ Reg. Conf. on Math. Phys.,
Trakya University, Edirne, Turkey (1991). }

\vskip .1cm  \item{[B.S]}{ V.V. Bazhanov and Yu.G. Stroganov,
{\sl Chiral Potts model as a
descendent of the six vertex model.} J. Stat. Phys. {\bf 51} (1990) 799.}

\vskip .1cm  \item{[D-C.K]}{  C. De Concini and V.G. Kac,
{\sl Representations of quantum groups at roots of 1.}
Progress in Math. {\bf 92} (1990) 471 Birkh\"auser,
and {\sl representations of quantum groups at roots of 1: reduction to the
exceptional case,} preprint RIMS 792 (1991).}

\vskip .1cm  \item{[D-C.K.P]}{  C. De Concini, V.G. Kac and C. Procesi,
{\sl Quantum coadjoint action.}
Preprint Pisa (1991).}

\vskip .1cm  \item{[D.J.M.M]}{ E. Date, M. Jimbo, K. Miki and
T. Miwa,
{\sl New $\cR$-matrices associated with cyclic representations of
$\cU_q(A_2^{(2)})$}, preprint RIMS-706 and
{\sl Generalized chiral Potts models and minimal cyclic
representations of
$\cU_q \left(\widehat{gl}(n,\CC )\right) $,}
Commun. Math. Phys. {\bf 137}  (1991) 133.}

\vskip .1cm  \item{[D.J.M.M.2]}{ E. Date, M. Jimbo, K. Miki and
T. Miwa,
{\sl Cyclic
representations of $\cU _{q}(sl(n+1,\CC ))$ at $q^{N}=1$,} Preprint RIMS-703
(1990), Publ. RIMS.
}

\vskip .1cm  \item{[Dob]}{ V. K. Dobrev, {\sl Proc. Int. Group Theory
Conference, St Andrews, 1989,} Vol. 1, Campbell and Robertson (eds.), London
Math. Soc. Lect. Notes Series 159, Cambridge University Press, 1991.}

\vskip .1cm  \item{[Dri]}{ V.G. Drinfeld, {\sl Quantum Groups,} Proc.
Int. Congress of Mathematicians, Berkeley, California, Vol.
{\bf 1}, Academic Press, New York (1986), 798.}

\vskip .1cm  \item{[F.G.P]}{ P. Furlan, A. C. Ganchev and V. B. Petkova,
{\sl Quantum groups and fusion rules multiplicities,}
Nucl. Phys. {\bf B343} (1990) 205.}

\vskip .1cm  \item{[F.R.T]}{ L. D. Faddeev, N.Yu. Reshetikhin and L.A.
Takhtajan, {\sl
Quantization of Lie groups and Lie algebras}, Leningrad Math.
J.  {\bf 1} (1990) No 1.}

\vskip .1cm  \item{[G.R]}{ G. von Gehlen and V. Rittenberg, {\sl
$\ZZ_n$-symmetric quantum chains with an infinite set of
conserved charges and
$\ZZ_n$ zero modes,} Nucl. Phys. {\bf B257} (1985) 351-370.}

\vskip .1cm  \item{[G.R-A.S]}{ C. G\'omez, M. Ruiz-Altaba and G. Sierra,
{\sl New $\cR$-matrix associated with finite dimensional representations
of $\cU_{q}(SL(2))$ at roots of unity,} Phys. Lett. {\bf B265} (1991) 95.}

\vskip .1cm  \item{[G.S]}{ C. G\'omez and G. Sierra,
{\sl A new solution  of the Star-Triangle equation based on $\cU _q(sl(2))$
at roots of unit,} Preprint CERN-TH.6200/91 and Gen\`eve  UGVA-DPT
1991/08-739.}

\vskip .1cm  \item{[H.R]}{ H. Hinrichsen and V. Rittenberg, {\sl A
two-parameter deformation of the $SU(1/1)$ superalgebra and the XY quantum
chain in a magnetic field,}
Preprint CERN-TH.6299/91, To appear in Phys. Lett.
{\bf B}, and references therein.}

\vskip .1cm  \item{[Jim]}{ M. Jimbo, {\sl $q$-difference analogue of
$\cU (\cG )$ and the Yang Baxter equation,} Lett. Math. Phys. {\bf 10} (1985)
63.}

\vskip .1cm  \item{[Kel]} { G. Keller,
{\sl Fusion rules of $\cU_{q}(SL(2,\CC))$, $q^{m}=1$.} Letters in
Math. Phys. {\bf 21} (1991) 273.}

\vskip .1cm  \item{[K.M.S]}{ R.M. Kashaev, V.V. Mangazeev and  Yu.G.
Stroganov,  {\sl Cyclic eight-state $\cR$-matrix related to
$\cU_q(sl(3))$ algebra at $q^{2}=-1$}, Preprint IHEP (1991), and
{\sl $N^3$-state $\cR$-matrix related with $\cU _q(sl(3))$ algebra at
$q^{2N}=1$}, preprint RIMS-823 (1991).}

\vskip .1cm  \item{[Lus]} { G. Lusztig,
{\sl Quantum groups at roots of $1$,} Geom. Ded. (1990).}

\vskip .1cm  \item{[M.R]}{ G. Moore and N. Reshetikhin, {\sl A comment on
quantum group symmetry in conformal field theory,} Nucl. Phys.  {\bf B328}
(1989) 557.}

\vskip .1cm  \item{[Nijs]}{ M. den Nijs, {\sl The domain wall theory of
two-dimensional commensurate-incommensu\-rate phase transitions,}
in {\sl Phase
transitions and critical phenomena, Vol. 12} C. Domb and J.L.
Lebowitz (eds.), Academic Press, New York (1988).}

\vskip .1cm  \item{[Pas]}{ V. Pasquier, {\sl Etiology of IRF models,}
Commun.
Math. Phys. {\bf 118}  (1988) 355.}

\vskip .1cm  \item{[P.S]}{ V. Pasquier and  H. Saleur, {\sl Common
structure between finite systems and conformal field theories through
quantum groups,} Nucl. Phys. {\bf B330}  (1990) 523.}

\vskip .1cm  \item{[R.A]}{  P. Roche and  D. Arnaudon,
{\sl Irreducible representations of the quantum
analogue of $SU(2)$.}
Lett. Math. Phys. {\bf 17} (1989) 295.}

\vskip .1cm  \item{[Ros]}{ M. Rosso, {\sl An analogue of the P. B. W.
theorem and the universal $\cR$-matrix for  $\cU _h sl(N+1)$,}
Commun. Math. Phys. {\bf 124} (1989) 307.}

\vskip .1cm  \item{[Skl]}{ E. K. Sklyanin, {\sl Some algebraic
structures connected with the
Yang-Baxter equation. Representations of quantum algebras,}
Funct. Anal. Appl. {\bf 17} (1983) 273.
}

\bye